\documentclass[draft,grl]{agu2001}
\usepackage{agu-ps}
\usepackage[dvips,final]{graphicx}
\usepackage{amssymb}
\ifgalley
\setlastpagenum{4}
\fi
\newcommand{\be}{\begin{equation}}
\newcommand{\ee}{\end{equation}}

\newcommand{\eqref}[1]{eq.\,(\ref{#1})}

\newcommand{\figwidth}{\hsize}

\authorrunninghead{COSTA} 
\titlerunninghead{CRYSTAL-CONTENT VISCOSITY DEPENDENCE}

\journalid{}

\articleid{}{}

\cpright{agu}{2005}

\revised{}
\accepted{}
\published{}

\authoraddr{A. Costa,
Department of Earth Sciences, University of Bristol, Wills Memorial
Building, Queen's Road, Bristol BS8 1RJ - UK. 
(e-mail: \hbox{a.costa@}bris.ac.uk)}
\slugcomment{Submitted in the {\it Geophysical Research Letters}, 2005.}

\begin{document}
\begin{figure}
\vspace{-2cm}
\centerline{\includegraphics[angle=0,height=0.8\figwidth]{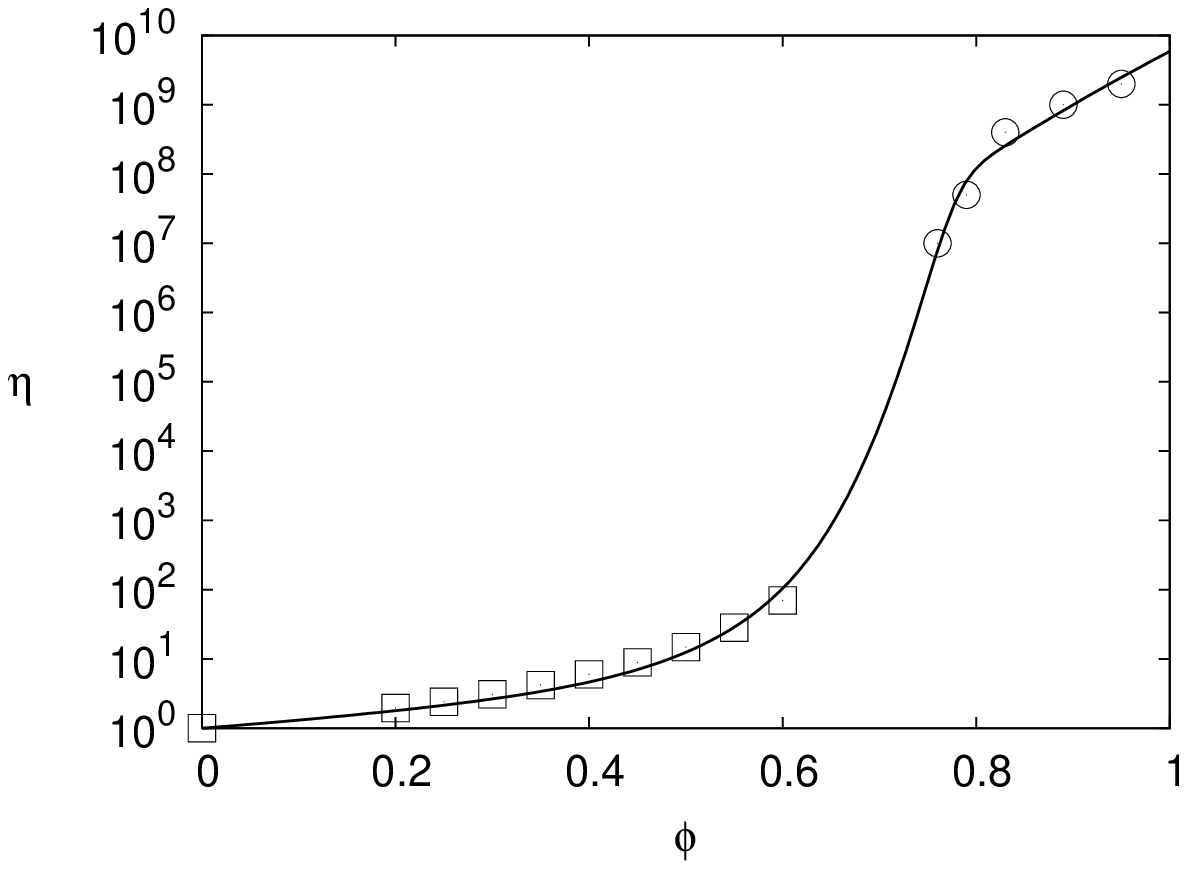}}
\caption{Effective relative viscosities deduced from data
by \citet{molpat79} at high solid fraction (circles),  and from
mean value of relative viscosities reported by \citet{tho65} at low
solid fraction (squares). For data by \citet{molpat79}, for
crystal-free fluid viscosity at 800$^o$C and 300 MPa, we considered a
value of $\mu_l=10^{4}$\,Pa\,s. The correlation coefficient is 0.999.
The best fit parameters are $\alpha=0.999916$, $\phi_*=0.673$ and
$\gamma=3.98937$, $\delta=16.9386$.}       
\label{total}
\end{figure}
\begin{figure}
\centerline{\includegraphics[angle=0,width=0.51\figwidth]{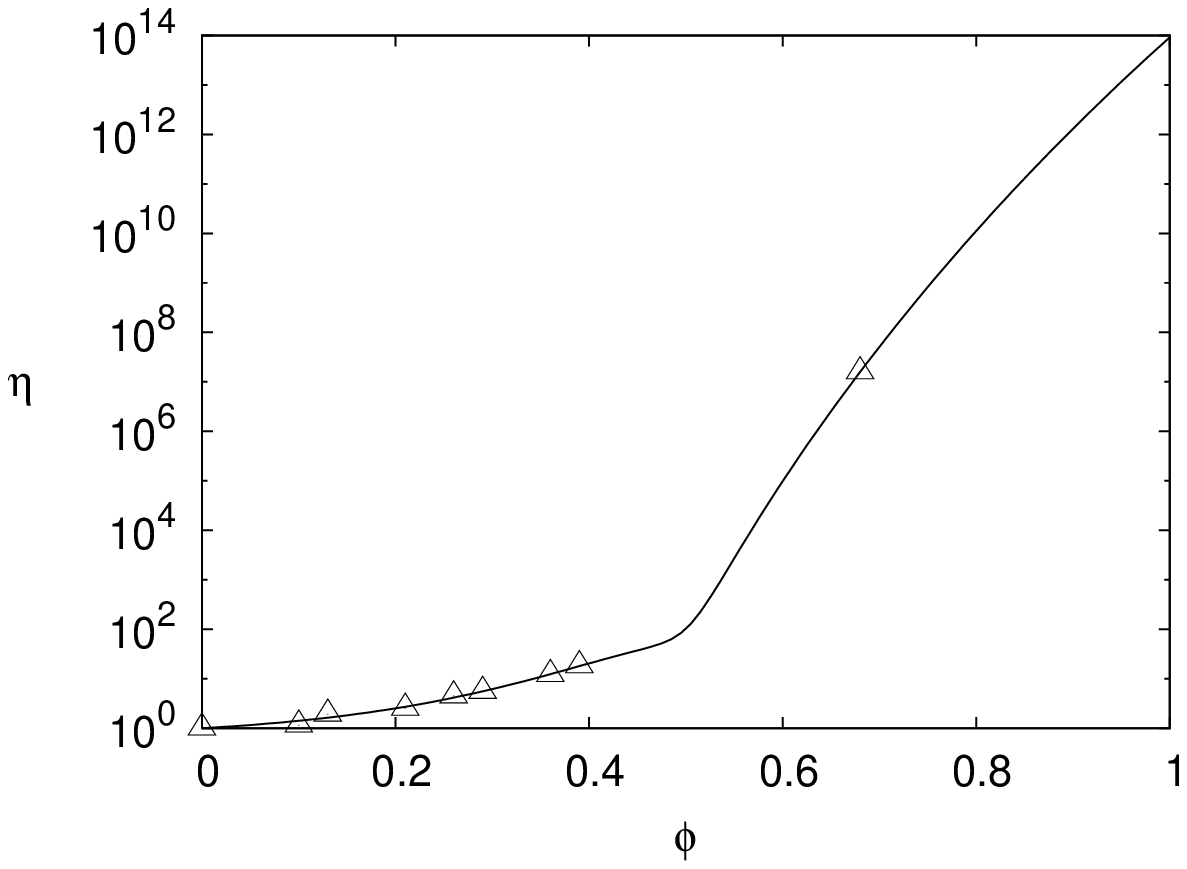}
\includegraphics[angle=0,width=0.51\figwidth]{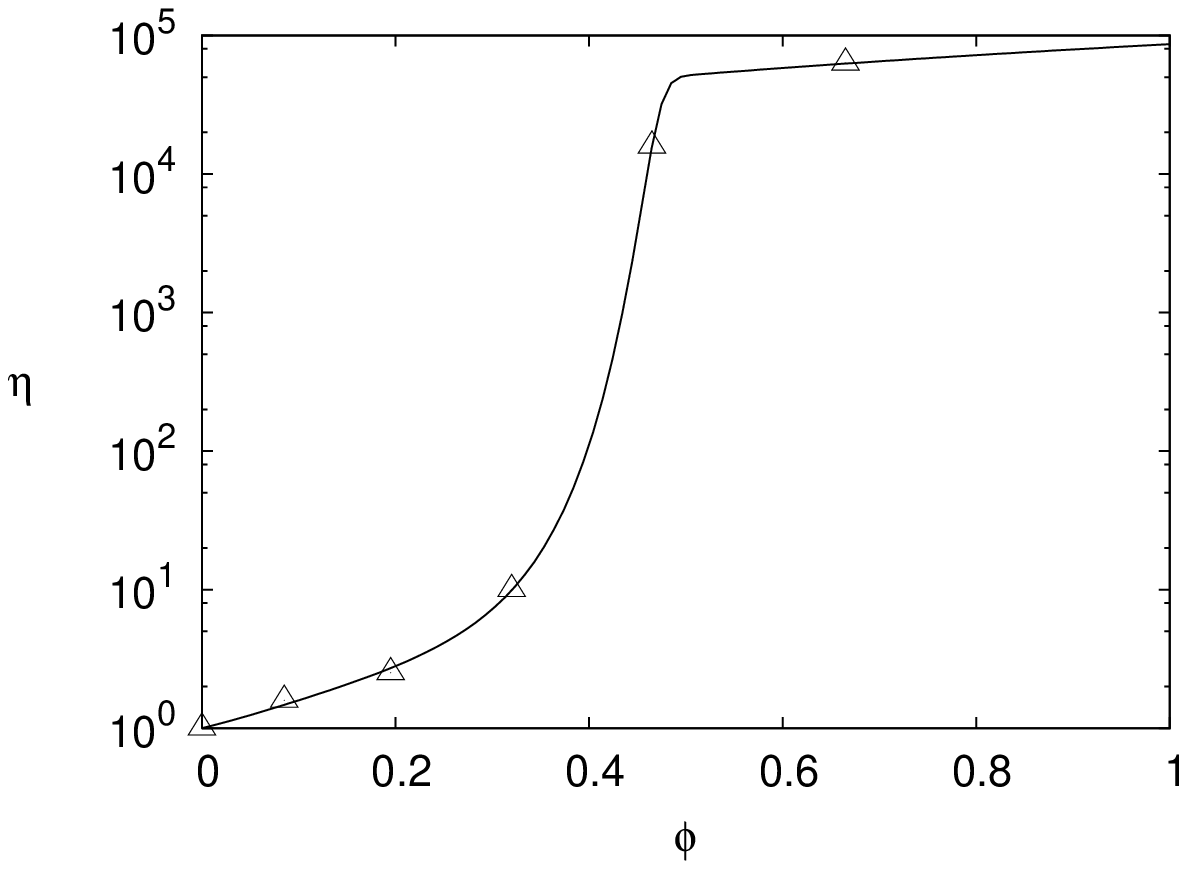}}
\caption{Relative viscosities from data 
  by \citet{lejric95} (triangles) and values using eq.~(\ref{costa2})
  (line) for Mg$_3$Al$_2$Si$_3$O$_{12}$(left) and Li$_2$Si$_2$O$_5$
  (right). The best fit parameters are $\alpha=0.9703;\, 0.9995\,$,
  $\phi_*=0.503;\,0.409$ and ${\gamma=0.924;\,4.214}$,
  ${\delta=40.386;\,1.149}$for Mg$_3$Al$_2$Si$_3$O$_{12}$ and
  Li$_2$Si$_2$O$_5$ respectively.}         
\label{lejric98}
\end{figure}

\title{Comments on ``Viscosity of high crystal content melts: dependence
  on solid fraction''} 
\author{Antonio Costa$^{1,2}$}
\affil{$^1$Centre for Environmental and Geophysical Flows, Department of Earth

Sciences, University of Bristol}
\affil{$^2$Also Istituto Nazionale di Geofisica e Vulcanologia, sezione
  di Napoli, Italy}
%
%
\begin{article}
%
%
A new parameterisation describing the relationship between viscosity
and solid fraction valid at large solid content was recently proposed
in \citet{cos2005}: 
\be
\eta(\phi)=\displaystyle\frac{1}{\left\{1- \alpha\,\mbox{erf}
\left(\displaystyle\frac{\sqrt{\pi}}{2}\phi\left[1+\displaystyle
\frac{\beta}{(1-\phi)^\gamma}\right]
\right)\right\}^{B/\alpha}}
\label{costa}
\ee
where $\eta$ is the relative viscosity, $\phi$ the volume fraction of
particles, $B$ the Einstein coefficient (with a theoretical value
$B=2.5$), and $\alpha,\beta,\gamma$ are three adjustable parameters. 
The proposed  parameterisation approximately reduces to the
classical, well established relationships for small fractions. On the
other hand, a weak point of equation~(\ref{costa}) is that for large
$\phi$, the relative viscosity tends quickly to a constant value,
because the non-linear term in the erf function rapidly saturates it
as $\phi$ approaches unity. \\
For very large solid fractions, no reliable experimental data are
available and there are also some intrinsic problems even in properly
defining relative viscosity \citep[see][]{cos2005}. However, we can
reasonably assume that the trend of the viscosity-crystal content
relationship obtained from data by \citet{molpat79} at high solid
fractions for partially-melted granite is generally valid, and this it
can be roughly described by a power law relationship. In fact, there
is no reason to assume that effective relative viscosity of melted
rocks tends towards a plateau region as $\phi\rightarrow 1$ as
predicted by (\ref{costa}). From a quantitative point of view it is no
simple to determine correctly the controlling parameters, such as the
critical fraction at which a rheological transition occurs or the
maximum value that effective relative viscosity can reach as
$\phi\rightarrow 1$. Moreover there are several indications that these
values strongly depend on the particle shape and particle size
distribution. However, since magma viscosity controls magma transport,
when modelling volcanic processes such as dome growth there is a
strong need to estimate the viscosity dependence on crystal-content,
even in the limit of very high crystal content where no experimental
observations are available \citep[see e.g.,][]{melspa2005}. \\
Here we propose a simple modification of relationship~(\ref{costa})
which improves all the main positive features of it, yet at the same 
time, does not show a plateau region as $\phi\rightarrow 1$, generally 
increases the performance of the model, and exactly recovers the
classical well established relationships valid for small fractions. In
order to obtain a satisfactory description of the effective relative
viscosity on the entire range of $\phi$, from zero to near the unity,
one more parameter is necessary, i.e. four parameters. The final four
parameter model we adopt is the following:     
\be
\eta(\phi)=\displaystyle\frac{1+\left(\displaystyle
\frac{\phi}{\phi_*}\right)^\delta}
{\left(1- \alpha\,\mbox{erf}
\left\{\displaystyle\frac{\sqrt{\pi}}{2\alpha} 
\displaystyle\frac{\phi}{\phi_*}
\left[1+\left(\displaystyle\frac{\phi}{\phi_*}\right)^\gamma\right]
\right\}\right)^{B\phi_*}}
\label{costa2}
\ee
where $0<\alpha<1$, $\phi_*$ represents the critical transition
fraction, $\gamma>1$ is a measure of the rapidity of the rheological
transition, and $\delta$ controls the increase of $\eta$ as
$\phi\rightarrow 1$. We can note that, for large $\delta$, the second
term at the numerator represents a negligible correction when $\phi
<\phi_*$, whilst it becomes important when $\phi >\phi_*$.  
It easy to see that as $\phi<\phi_*$ with decreasing $\phi$, the
parameterisation~(\ref{costa2}) tends exactly to the \citet{kridou59} 
relationship:    
\be
\eta= \left(1-\frac{\phi}{\phi_*}\right)^{-B\phi_*}
\label{roscoe3}
\ee
and as $\phi\rightarrow 0$, it recovers exactly the Einstein
equation:    
\be
\eta(\phi)\simeq\left(1+B\phi\right)
\label{einstein}
\ee
where $B$ is the Einstein coefficient. \\
An example of the good performance of relationship~(\ref{costa2}) is
shown in Figure~\ref{total}. Here we plotted relative viscosities
deduced from the data of \citet{molpat79} in the high crystal fraction
regime, considering a crystal-free fluid viscosity of $10^{4}$\,Pa\,s
as suggested by \citet{molpat79}. These values are matched with and
mean values of relative viscosities reported by
\citet{tho65}. Although we must keep in mind all the intrinsic
limitations of these data \citep[see][]{cos2005}, we can observe that
over the entire range of solid fraction, the model is able to 
reproduce all data with an excellent correlation ($R^2=0.999$) and the
curve behaviour suggested by \citet{molpat79}.    
Another example is shown in Figure~\ref{lejric98} where, as in
\citet{cos2005}, the relative viscosities for
Mg$_3$Al$_2$Si$_3$O$_{12}$ and Li$_2$Si$_2$O$_5$ measured by
\citet{lejric95} and the fitted values are shown. However a direct
comparison with Figure~2 of \citet{cos2005} is not possible because
the relative viscosities values reported in Figure~2 of
\citet{cos2005} were erroneously underestimated in systematic way for
both Mg$_3$Al$_2$Si$_3$O$_{12}$ and Li$_2$Si$_2$O$_5$. 
%
%
\acknowledgement
This work was supported by NERC research grant reference
NE/C509958/1. The author would like to thank O. Melnik for his
useful comments and for the discussions had during his stay at the
Institute of Mechanics, Moscow State University, in December 2005.
%
%
\bibliography{references}
\end{article}
\end{document}